\documentclass{aastex63}

\usepackage{color}

\graphicspath{{./}{figures/}}

\begin{document}

\title{Multiple Rings and Asymmetric Structures in the Disk of SR 21}

\correspondingauthor{Yi Yang}
\email{yi.yang@abc-nins.jp}

\author[0000-0002-9024-4150]{Yi YANG}
\affiliation{National Astronomical Observatory of Japan (NAOJ), National Institutes of Natural Sciences (NINS), 2-21-1 Osawa, Mitaka, Tokyo 181-8588, Japan}
\affiliation{Department of Astronomy, The University of Tokyo, 7-3-1 Hongo, Bunkyo-ku, Tokyo 113-0033, Japan}
\affiliation{Astrobiology Center, NINS, 2-21-1, Osawa, Mitaka, Tokyo 181-8588, Japan}

\author[0000-0003-2300-2626]{Hauyu Baobab LIU}
\affiliation{Academia Sinica Institute of Astronomy and Astrophysics, P.O. Box 23-141, Taipei 10617, Taiwan}

\author{Takayuki MUTO}
\affiliation{Division of Liberal Arts, Kogakuin University, 1-24-2 Nishi-Shinjyuku, Shinjyuku-ku, Tokyo 163-8677, Japan}

\author{Jun HASHIMOTO}
\affiliation{Astrobiology Center, NINS, 2-21-1, Osawa, Mitaka, Tokyo 181-8588, Japan}
\affiliation{National Astronomical Observatory of Japan (NAOJ), National Institutes of Natural Sciences (NINS), 2-21-1 Osawa, Mitaka, Tokyo 181-8588, Japan}

\author{Ruobing DONG}
\affiliation{Department of Physics and Astronomy, University of Victoria, 3800 Finnerty Road, Victoria, BC V8P 5C2, Canada}

\author{Kazuhiro KANAGAWA}
\affiliation{College of Science, Ibaraki University, 2-1-1 Bunkyo, Mito, Ibaraki 310-8512, Japan}

\author[0000-0002-3001-0897]{Munetake MOMOSE}
\affiliation{College of Science, Ibaraki University, 2-1-1 Bunkyo, Mito, Ibaraki 310-8512, Japan}

\author[0000-0002-5082-8880]{Eiji Akiyama}
\affiliation{Division of Fundamental Education and Liberal Arts, Department of Engineering, Niigata Institute of Technology 1719 Fujihashi,
Kashiwazaki, Niigata 945-1195, Japan}

\author[0000-0002-9017-3663]{Yasuhiro HASEGAWA}
\affiliation{Jet Propulsion Laboratory, California Institute of Technology, Pasadena, CA 91109, USA}

\author[0000-0002-6034-2892]{Takashi TSUKAGOSHI}
\affiliation{Faculty of Engineering, Ashikaga University, Ohmae-cho 268-1, Ashikaga, Tochigi, 326-8558, Japan}

\author[0000-0003-0114-0542]{Mihoko KONISHI}
\affiliation{Faculty of Science and Technology, Oita University, 700 Dannoharu, Oita 870-1192, Japan}

\author{Motohide TAMURA}
\affiliation{Department of Astronomy, The University of Tokyo, 7-3-1 Hongo, Bunkyo-ku, Tokyo 113-0033, Japan}
\affiliation{Astrobiology Center, NINS, 2-21-1, Osawa, Mitaka, Tokyo 181-8588, Japan}
\affiliation{National Astronomical Observatory of Japan (NAOJ), National Institutes of Natural Sciences (NINS), 2-21-1 Osawa, Mitaka, Tokyo 181-8588, Japan}

\begin{abstract}

Crescent-like asymmetric dust structures discovered in protoplanetary disks indicate dust aggregations. Thus, the research on them helps us understand the planet formation process. Here we analyze the ALMA data of the protoplanetary disk around the T-Tauri star SR 21, which has asymmetric structures detected in previous sub-millimeter observations. Imaged at ALMA Band 6 (1.3 mm) with a spatial resolution of about 0.$\arcsec$04, the disk is found to consist of two rings and three asymmetric structures, with two of the asymmetric structures being in the same ring. 
Compared to the Band 6 image, the Band 3 (2.7 mm) image also shows the three asymmetric structures but with some clumps. The elongated asymmetric structures in the outer ring could be due to the interactions of a growing planet.
By fitting the Band 3 and Band 6 dust continuum data, two branches of solutions of maximum dust size in the disk are suggested: one is larger than 1 mm, and the other is smaller than 300 $\mu m$. High-resolution continuum observations at longer wavelengths as well as polarization observations can help break the degeneracy. We also suggest that the prominent spiral previously identified in VLT/SPHERE observations to the south of the star at 0.$\arcsec$25 may be the scattered light counterpart of the Inner Arc, and the structure is a dust-trapping vortex in nature. The discovered features in SR 21 make it a good target for studying the evolution of asymmetric structures and planet formation.

\end{abstract}

\keywords{planets and satellites: formation --- protoplanetary disks: SR 21 ---
planet-disk interactions}

\section{Introduction} \label{sec:Introductions}
In recent years high angular resolution observations in the sub-millimeter band using instruments such as the Atacama Large Millimeter/submillimeter Array (ALMA) have discovered that some protoplanetary disks show asymmetric structures, such as HD 142527 \citep[e.g.,][]{Fukagawa2013-da,Casassus2013-nm,Boehler2017}, Oph IRS 48 \citep{VanderMarel2013}, SAO 206462 (HD 135344B) \citep[e.g.,][]{Perez2014,Cazzoletti2018-gi} and MWC 758 \citep[e.g.,][]{Dong2018-ve}. \citet{Francis2020-jv} discovered that in a sample of 38 transition disks, the fraction of asymmetric disks is 24\%, while the occurrence rate of asymmetric structures in all kinds of disks remains unknown. It is believed that dust is "trapped" in these structures due to pressure bumps in the disk, which accelerate dust growth and promote planet formation processes \citep[e.g.,][]{Barge1995-nw,Klahr1997-qv,Regaly2012-zz}. Therefore, research on these asymmetric structures in the protoplanetary disks is vital for us to learn about the planet formation process in protoplanetary disks.

Several scenarios have been suggested for the formation of asymmetric structures. One popular scenario suggests that such asymmetry is caused by Rossby Wave Instability (RWI) vortex \citep[e.g.,][]{Zhu2014}. Under a low viscosity ($\alpha\sim10^{-5}-10^{-4}$), the vortices, as well as the asymmetric structures, can be sustained for a few Myrs \citep{Fu2014-zt}. Besides, other scenarios, such as an eccentric cavity \citep[e.g.,][]{Ataiee2013} and "horseshoe structures" \citep[e.g.,][]{Ragusa2017-is}, could also be responsible for these structures. Understanding the formation mechanisms of such asymmetric structures will no doubt improve our understanding of planet formation and evolution.

It is believed that the asymmetric structures will finally dissipate, and they elongate and break with the interaction of a growing planet \citep[e.g.,][]{Hammer2019} or due to the dust back reaction onto gas \citep{Fu2014-zt}. Above all, we still know little about how asymmetric structures evolve and how the planets form from them. Observing disks with asymmetric structures, and trying to understand their dust sizes, is quite necessary for constraining theoretical models.

Asymmetric structures are also related to other asymmetric structures. They could appear in multiple-ring systems (e.g., SAO 206462 \citep[e.g.,][]{Perez2014,Cazzoletti2018-gi}, HD 143006 \citep{Perez2018-rm}), and \citet{Garufi2018-ew} found out that all asymmetric disks seen in ALMA have spiral arms in the near-infrared band. \citet{Van_der_Marel2021-vr} suggested that asymmetric dust structures could be linked to a low local gas surface density through the observational Stokes number, which can explain why the asymmetric features in multi-ring systems tend to appear in the outer ring, such as SAO 206462 and HD 143006, and can also be linked to either vortices (for $\alpha\lesssim10^{-4}$ in disks) or spiral arms.

SR 21 is a T-Tauri star located in the Ophiuchus Molecular Cloud at a distance of about 138.4$\pm$1.1 pc from Earth \citep{GAIA2018}. \citet{Herczeg2014-dm} suggested that this F7 type star has a stellar mass of about 1.67 $M_\odot$, luminosity of about 7.4 $L_\odot$ and an age of about 10 Myr through a low-resolution spectroscopic study. Previous near-infrared and sub-millimeter observations \citep[e.g.,][]{Andrews2011,Follette2013,Muro-Arena2020} have shown that it holds a circumstellar disk, which has a cavity in sub-millimeter continuum emission but not in near-infrared scattered light, suggesting segregation of dust of different sizes \citep{dong12}. \citet{Perez2014} revealed that it has an asymmetric structure in submillimeter observations at a spatial resolution of about 0.$\arcsec$3 using ALMA Band 9 (0.45 mm). \citet{Perez2014} also pointed out that its inclination is about 15$^\circ$ and position angle is about 16$^\circ$, so it is a nearly face-on disk and suitable for asymmetric structure study. Near Infrared observations using the Spectro-Polarimetric High-contrast Exoplanet REsearch \citep[SPHERE, ][]{Beuzit2019-um} mounted on the Very Large Telescope (VLT) showed that it has complex structures including rings and spiral arms, under a spatial resolution of about 0.05$\arcsec$ \citep{Muro-Arena2020}. \citet{Muro-Arena2020} also pointed out that the spiral arms of SR 21 are inside an outer ring and quite tight (pitch angles close to zero at a radius of $\sim0.\arcsec2$), while the spiral arms in other targets like HD 135344B and MWC 758 are the outer-most visible structures and relatively loose (pitch angles$\sim10^\circ-16^\circ$). While previous observations of SR 21 in sub-millimeter bands have spatial resolutions of only about 0.2$\arcsec$-0.3$\arcsec$, it is not enough to understand the corresponding structures of large dust in the disk. Therefore, higher-resolution observations in the sub-millimeter band are pretty necessary.

This paper will show our new observational results of SR 21 via ALMA at a higher resolution. In Section 2, we will describe our observations; in section 3, we will explain our results; in section 4, we will give some discussions by combining our observation results and previous results; and in section 5, we will make a summary.

\section{Observations and Data Reduction} \label{sec:Observations}

We have carried out the ALMA observations towards SR 21 (2018.1.00689.S, PI: T. Muto) at Band 6 in 2019, in the C43-6 (short-spacing; August 24; $\sim$10 minutes on source) and C43-9 (long baseline; June 24; $\sim$48 minutes on source) array configurations.
The C43-6 and C43-9 configuration observations covered the baseline ranges of $\sim$40--3400 and $\sim$80--13640 meters, respectively.
The average precipitable water vapor (PWV) during the C43-6 and C43-9 observations were 1.0 and 0.5 mm, respectively.

Our correlator setup employed five spectral windows: there were two continuum spectral windows (1.875 GHz bandwidth, 43 km\,s$^{-1}$ velocity resolution) centering at the rest frequencies of 217 GHz and 233 GHz; two spectral windows were centering at the rest frequencies of $^{13}$CO 2-1 (220.398684 GHz) and C$^{18}$O 2-1 (219.560358 GHz) of which the bandwidth and velocity resolution are 468.75 MHz and $\sim$0.665 km\,s$^{-1}$, respectively; there was one spectral window centering at the request frequency of CO 2-1 (230.538000 GHz) of which the bandwidth and velocity resolution are 937.50 MHz and $\sim$0.635 km\,s$^{-1}$, respectively.

The basic calibration was carried out using the pipeline in the Common Astronomy Software Applications (CASA) package 5.4.0-70.
We derived the Bandpass and complex gain calibration solutions based on the observations on J1517-2422 and J1633-2557, respectively.
The absolute flux scaling was derived based on referencing the ALMA calibrator grid survey on J1517-2422.
We quote the proper motion of the SR 21 from GAIA DR2 \citep{GAIA2018}.
To allow joint imaging of all data, we used the CASA task \texttt{fixplanets} to shift the target source to the expected coordinates on June 23, 2019.

We used the CASA task \texttt{uvcontsub} to separate the continuum and spectral line data.
We perform phase self-calibration on the continuum data only and then apply these solutions to the line emission.
We combined all spectral windows when deriving self-calibration solutions.
In addition, we combined the two polarizations (i.e., XX and YY) for the long baseline observations to avoid massive flagging due to the limited signal-to-noise ratios.
The solution intervals for the C43-6 and C43-9 configuration observations were 30 seconds and 540 seconds, respectively.
The longer solution interval for the C43-9 configuration observations was due to the lower signal-to-noise ratios. The command \texttt{tclean} provided by CASA was used to image the continuum data, with Briggs weighting and robust parameter 0.5. While we used the \textsc{Miriad} software package \citep{Sault1995} to perform naturally weighted imaging for the CO 2-1, $^{13}$CO 2-1, and C$^{18}$O 2-1 lines.

\section{Results} \label{sec:Results}

\subsection{ALMA Band 6 Continuum image}

The Band 6 continuum image of SR 21 is shown in Figure~\ref{fig:fig1}(a). The beam size is $0.\arcsec050\times0.\arcsec047$ with a position angle 94$^\circ$. From the image, one can see two rings. The outer ring ($>25\sigma$, 1$\sigma\sim1.7\times10^{-5}$Jy/beam) extends between a radial separation from the central star of 0.34$\arcsec$ to 0.48$\arcsec$ (47-66 au). As for the inner ring, it extends between a radial separation of 0.14$\arcsec$ to 0.28$\arcsec$ (19.5-39 au), the faintest part on the north side reaches $5\sigma$ while its brightest part on the south side is higher than $\sim20\sigma$.

Asymmetric structures can be found in the inner and outer rings.
The outer ring has two asymmetric structures: South Arc (peak located at $\sim170^\circ$) and North Arc (peak located at $\sim305^\circ$). While for the asymmetric structures of the inner ring, we refer to it as the Inner Arc (peak located at $\sim165^\circ$). We notice that actually, the Inner Arc's radius is only from 0.$\arcsec$18 to 0.$\arcsec$28, and the ring structure between 0.$\arcsec$14 to 0.$\arcsec$18 near the Inner Arc is significantly fainter, which may indicate that the inner ring consists of two "sub-rings": one standard ring and one ring with the Inner Arc. In this paper, we do not distinguish between them and simply regard them as the "inner ring".

\subsubsection{Comparison with ALMA Band 3 image}

\begin{figure}[htbp!]
\gridline{
    \fig{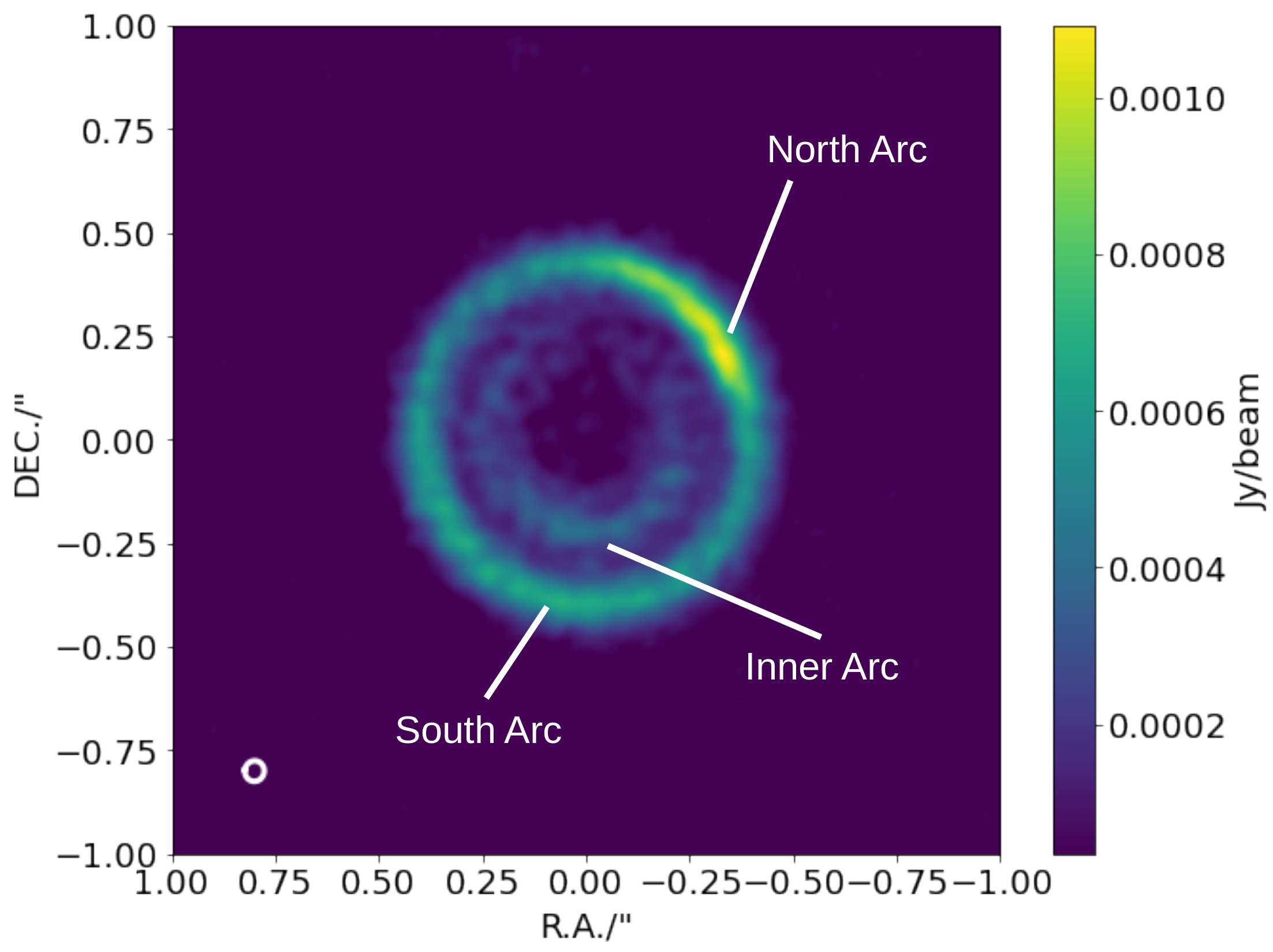}{0.45\textwidth}{a}
    \label{fig:fig1a}
    \fig{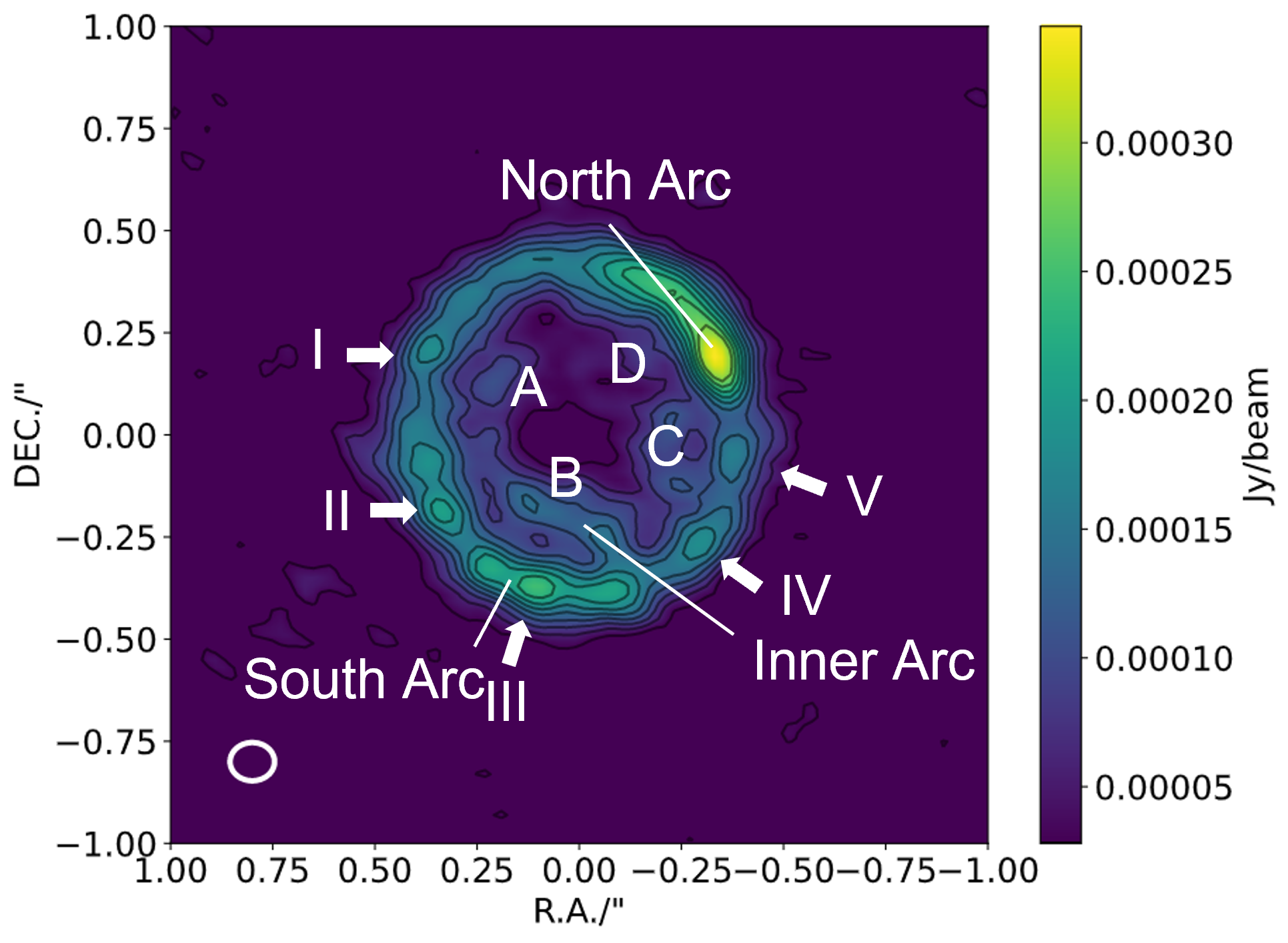}{0.45\textwidth}{b}
    \label{fig:fig1b}
	}
	\caption{(a): Band 6 image of SR 21. $1\sigma\sim1.7\times10^{-5}$Jy/beam; (b): Band 3 image of SR 21 with contours. Contour interval is $2\sigma$($1\sigma\sim1.4\times10^{-5}$Jy/beam).}
\label{fig:fig1}
\end{figure}

We compare our Band 6 observations with some previous observations. Firstly we compared our Band 6 image with ALMA Band 3 (107.01 - 110.67 GHz, 2.7-2.8 mm) continuum image. The Band 3 image (Project ID: 2017.1.00884.S, PI: Pinilla) shown in Figure~\ref{fig:fig1}(b) is the same as the image in \citet{Muro-Arena2020}, with the baseline ranging from 138 m to 13.8 km ( 50-5000 $k\lambda$). The inner and the outer ring are barely resolved under a beam size of 0.$\arcsec$10$\times$0.$\arcsec$09, and the South Arc and North Arc ($>20\sigma$, $1\sigma\sim1.4\times10^{-5}$Jy/beam) in the outer ring can also be seen. In Band 3, the outer ring has several small "clumps", which differs from the Band 6 image. Some notable clumps and their parameters are summarized in Table~\ref{table:table1}. 

\begin{deluxetable*}{cccccc}
\label{table:table1}
\tablecaption{Notable Clumps detected in Band 3 Image}
\tablewidth{0pt}
\tablehead{Outer Ring & I & II & III & IV & V }
\startdata
PA($^\circ$) & 60 & 120 & 165 & 225 & 265 \\
SNR($\sigma$) & 12 & 14 & 16 & 12 & 10 \\
\hline
Inner Ring & A & B1 & B2 & C & D \\
\hline
PA($^\circ$) & 65 & 145 & 195 & 245 & 320 \\
SNR($\sigma$) & 7 & 9 & 10 & 7 & 5 \\
\enddata
\end{deluxetable*}

\begin{figure}[htbp!]
\gridline{
    \fig{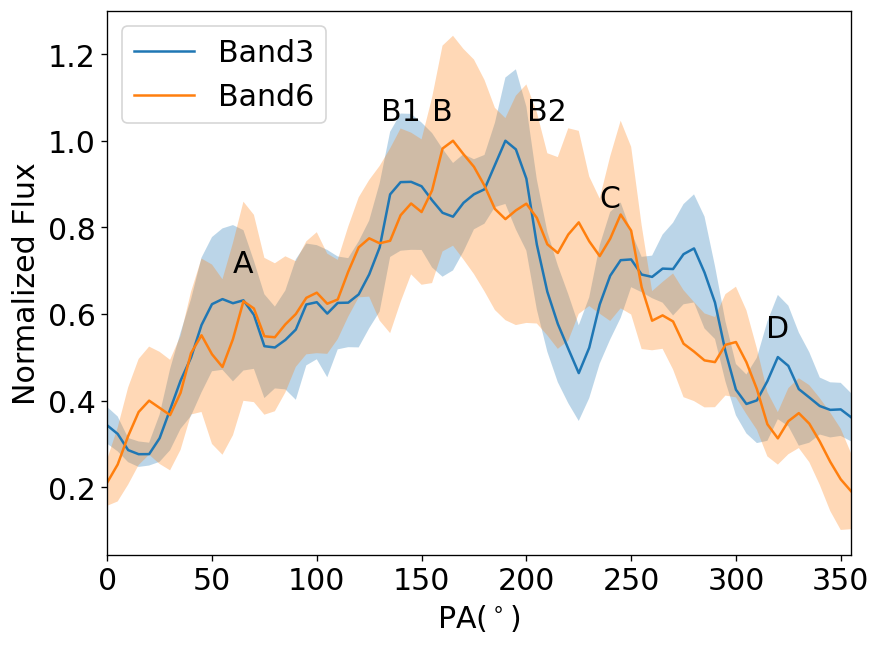}{0.45\textwidth}{a}
    \label{fig:fig3a}
    \fig{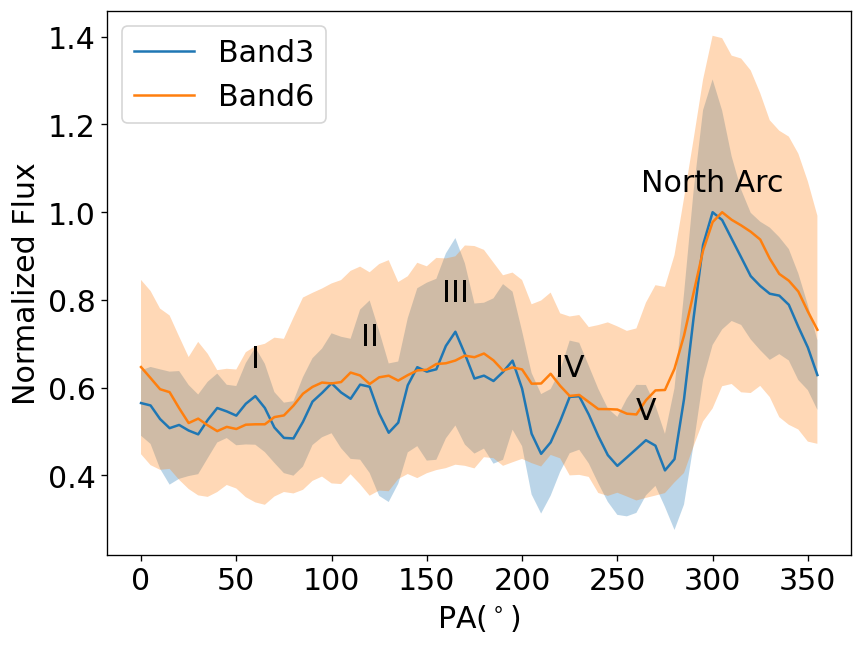}{0.45\textwidth}{b}
    \label{fig:fig3b}
	}
	\caption{Azimuthal profiles of SR 21 inner ring (a) and outer ring (b). The shaded regions indicate the uncertainties. The inner ring profile includes pixels between 0.14” and 0.28” from the central star, while the outer ring profile includes pixels between 0.34” to 0.48” from the central star. The profile is drawn by averaging over bins of 10$^\circ$ width, and the uncertainty is the standard deviation of every bin.}
\label{fig:fig3}
\end{figure}

To have a detailed comparison of the asymmetric structures in the two different bands, we draw the azimuthal profiles of the inner ring and the outer ring in Band 3 and Band 6 (Figure~\ref{fig:fig3}). The profile is drawn by averaging over bins of 10$^\circ$ widths. Figure~\ref{fig:fig3}(a) compares the azimuthal profile of the inner ring (0.14$\arcsec$-0.28$\arcsec$). For the Band 6 data, an obvious peak B is located at $165^\circ$ corresponding to the Inner Arc. While for Band 3 data, its highest peak ($>8\sigma$) corresponding to the Inner Arc has two sub-peaks, located at 145$^\circ$ and 195$^\circ$, respectively, so this peak centers at 170$^\circ$. Besides, there are three other obvious peaks, A ($\sim7\sigma$) and C($\sim7\sigma$), located at $\sim65^\circ$ and $\sim245^\circ$, and a relatively faint peak D ($\sim5\sigma$) detected at PA $\sim320^\circ$ in Band 3, but in Band 6 no corresponding structures can be found. 

The azimuthal profiles of the outer ring (0.$\arcsec$34-0.$\arcsec$48) are shown in Figure~\ref{fig:fig3}(b). Band 6 clearly shows two peaks, one ranges from 40$^\circ$ to 250$^\circ$, and its peak locates at 170$^\circ$, corresponding to the South Arc, while the other peak locates at 305$^\circ$, corresponding to the North Arc. From this figure, it can be seen that the profiles of the arcs are asymmetric: their peaks are not at their center positions of them. As for Band 3, its azimuthal profile shows five small peaks located at 60$^\circ$, 120$^\circ$, 165$^\circ$, 225$^\circ$, 265$^\circ$ corresponding to the clumps I-V mentioned above, and the highest of them locates at about $165^\circ$. As for the North Arc in Band 3, it locates at about $300^\circ$.

These clumps show higher contrasts at Band 3, which may be partly due to the lower optical depths at this Band. On the other hand, some of the clumps detected at $\lesssim$2-$\sigma$ have a 5\% chance to be spurious. To assess whether or not a clump may be spurious, we measured the brightness contrast between its peak and the "valleys" with the lowest intensity in the inter-clump regions. For the outer ring, I and V are only 2$\sigma$ brighter than their adjacent valleys; II is 4$\sigma$ brighter; IV is 2$\sigma$ brighter than the valley between III and IV but is 4$\sigma$ brighter than the valley between IV and V. In light of this, we consider II could be more robust than I, IV and V. As for the inner ring, D is only about 2$\sigma$ brighter; A is 4$\sigma$ brighter than the valley between A and D, but is only 2$\sigma$ brighter than the valley between A and B; C is 4$\sigma$ brighter than the valley between C and D, but only 2$\sigma$ brighter than the valley between C and B. The two sub-peaks of B are only about 1-2$\sigma$ brighter than the valley between them. So despite the North Arc and the South Arc (III) as well as the Inner Arc (B), which are indeed localized clumps since they are observed in both two Bands, we suggest that other clumps, I, IV, V, A, C, D, as well as the two sub-peaks of B, may be spurious, while II could represent the actual structure since it is 4$\sigma$ brighter than its nearby valleys. More profound observations at longer wavelength bands are required to confirm these clumps.

\subsubsection{Comparison with VLT/SPHERE H-Band image}

\begin{figure}[htbp!]
\gridline{
    \fig{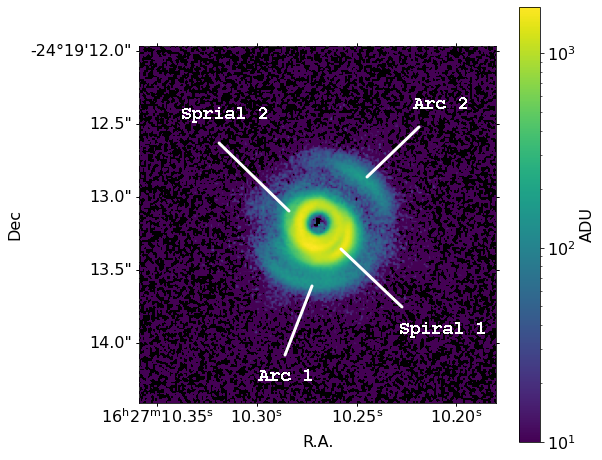}{0.5\textwidth}{a}
    \label{fig:fig4a}
    \fig{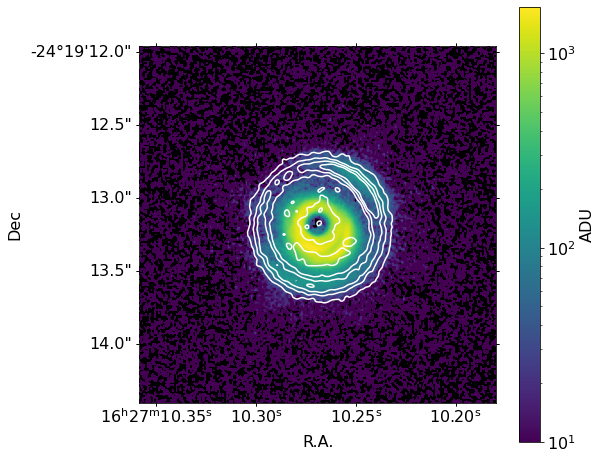}{0.5\textwidth}{b}
    \label{fig:fig4b}
	}
\caption{Near-infrared image of SR 21 taken by VLT/SPHERE/IRDIS. \textit{Left}: Near-infrared image with labels showing structures detected by \citet{Muro-Arena2020}; \textit{Right}: Near-infrared image with Band 6 continuum data (contour) overplotted. The contour intervals are 4$\sigma$ ($\sim$6.8$\times10^{-5}$ Jy/beam).}
\label{fig:fig4}
\end{figure}

We also compared our data with the previous H-Band ($\sim$1.6 $\mu$m) image taken by VLT/SPHERE. The SPHERE archival data were reduced by the IRDAP (IRDIS Data reduction for Accurate Polarimetry) pipeline \citep{2020A&A...633A..64V,2017SPIE10400E..15V}, the same as \citet{Muro-Arena2020}. We overplotted our Band 6 data on the SPHERE data, shown in Figure~\ref{fig:fig4}. From the image, it can be seen that for the inner ring, our Inner Arc is generally consistent with the Spiral 1 structure reported by \citet{Muro-Arena2020}, indicating some relationship between this asymmetric structure and the spiral arm. As for the outer ring, the South Arc fits well with Arc 1, while the North Arc corresponds to Arc 2.

As mentioned in Section 1, it is common that spiral arms seen in the near-infrared band accompanied by asymmetric structures in the sub-millimeter band, such as HD 135344B \citep{Cazzoletti2018-gi} and MWC 758 \citep{Dong2018-ve}. However, for SR 21, there are some differences. Despite the tight spiral arm \citet{Muro-Arena2020} has pointed out, we also noticed that the asymmetric structure of the Inner Arc corresponds to two spiral arms:  Spiral 1 and Spiral 2, and their positions fit quite well without obvious offset; this is also different from other targets: the asymmetric structures in HD 135344B and V1247 Ori \citep{Kraus2017-uf} only connect the tip of the spiral arms, while in HD 100546 \citep{Rosotti2020-rk} the asymmetric structure only locates at the start of the spiral arm, and in MWC 758, two asymmetric structures correspond to only one spiral arm. 

\subsection{ALMA Band 6 CO Isotopologue Emissions}

\begin{figure}
\plotone{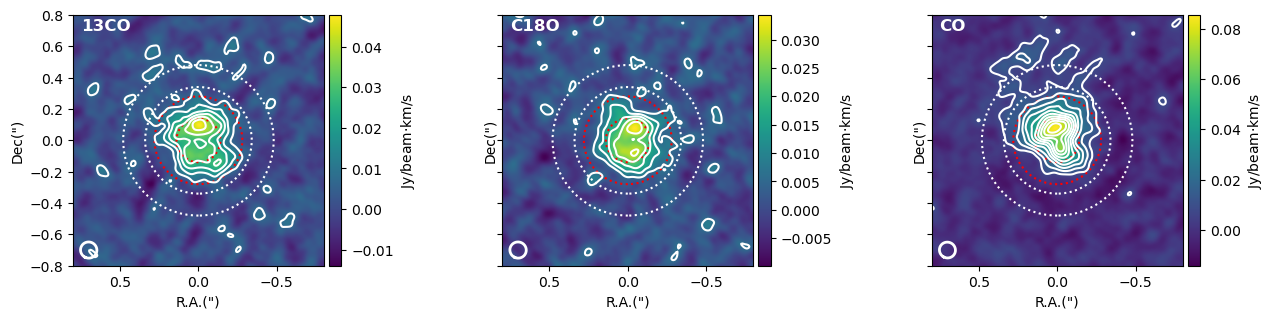}
\caption{CO 2-1 isotopologue emission images of SR 21. The white dotted lines indicate the outer ring, while the red dotted lines indicate the inner ring. The contour intervals are 2$\sigma$ (1$\sigma\sim$0.0035, 0.003, and 0.004 Jy/beam$\cdot$km/s for $^{13}$CO, C$^{18}$O and CO images, respectively).}
\label{fig:figco1}
\end{figure}

Three CO lines, $^{12}$CO, $^{13}$CO, and C$^{18}$O (2-1) were observed in our research.
Their beam sizes are $0.\arcsec103\times0.\arcsec100$ with PA=59$^\circ$, $0.\arcsec104\times0.\arcsec103$ with PA=64$^\circ$ and $0.\arcsec104\times0.\arcsec103$ with PA=65$^\circ$, respectively. Their spectral resolutions of them are 0.6 km/s. We made the moment 0 (intensity) maps of them using the command \texttt{immoments}, and they are shown in Figure~\ref{fig:figco1}. The start(end) of the channels used to make moment maps are chosen by checking that if a bright source ($>4\sigma, 1\sigma=0.001 Jy/beam$) within 0.5” from the central stars appears/disappears. In detail, for the $^{12}$CO line, the channels between -1.96km/s and 8.09 km/s are chosen, for $^{13}$CO line, the channels between -1.96km/s and 6.75 km/s are chosen, and for C$^{18}$O line the channels between -0.62km/s and 6.75 km/s are chosen. The root mean squared (RMS) noise levels of the moment 0 maps are 0.035, 0.03, and 0.04 Jy/beam for $^{13}$CO, C$^{18}$O, and $^{12}$CO images, respectively. Generally, the observed gas structures in protoplanetary disks should be more extended than observed dust structures due to different optical depths \citep[e.g.,][]{Facchini2017-hy}. However, in these figures, the CO features are quite compact, and the emissions mostly come from inside the inner ring, i.e., $<$0.$\arcsec$28. One exception is the $^{12}$CO 2-1 image, which shows an extended structure to about 0.$\arcsec$8 in the northeast, although it only has 2-3 $\sigma$.

The $^{13}$CO and C$^{18}$O images show two cores in the north and south. Their positions are slightly shifted in different wavelengths: for the $^{13}$CO image, they have position angles of about 0 and 210$^\circ$, and for the C$^{18}$O image, they have position angles of about -40 and 200$^\circ$. As for the $^{12}$CO image, only the north core (PA$\sim10^\circ$) is apparent.

The compact features in the CO isotopologue emissions seem to indicate the depletion of gas outside of $0.\arcsec28$. However, considering the large extinctions of SR 21 ($A_V$=6.2, \citep{Herczeg2014-dm}), one more reasonable explanation is that they are affected by the foreground adsorptions. The emission component can be approximated by a Gaussian curve since SR 21 seems close in face-on. To determine the strengths, velocity, and width of the emission and absorption lines, we fit the spectra with the following equation \citep{Kim2020-qi}:

\begin{equation}
    I_v=A_{emit}\mbox{exp}(-\frac{(V-V_{emis})^2}{2\Delta V_{emis}^2})+A_{abs}\mbox{exp}(-\frac{(V-V_{abs})^2}{2\Delta V_{abs}^2})
\end{equation}

\begin{deluxetable*}{ccccccc}
\label{tab:tab1}
\tablecaption{Fitting Results of the CO Isotopic Lines\label{tab:spectra}}
\tablewidth{0pt}
\tablehead{\colhead{Species} & \colhead{$A_{emis}$} & \colhead{$V_{emis}$} & \colhead{$\Delta V_{emis}$} &
\colhead{$A_{abs}$} & \colhead{$V_{abs}$} & \colhead{$\Delta V_{abs}$} \\
\colhead{} & \colhead{(mJy/beam)} & \colhead{(km/s)} & \colhead{(km/s)} &
\colhead{mJy/beam} & \colhead{(km/s)} & \colhead{(km/s)}}
\startdata
$^{12}$CO & 7.354$\pm$3.588 & 2.500$\pm$0.284 & 0.616$\pm$0.059 & -8.000$\pm$3.591 & 3.134$\pm$0.121 & 0.980$\pm$0.162 \\
$^{13}$CO & 4.065$\pm$0.893 & 2.841$\pm$0.055 & 0.626$\pm$0.040 & -4.346$\pm$0.870 & 3.039$\pm$0.032 & 1.144$\pm$0.096 \\
C$^{18}$O & 2.129$\pm$1.231 & 3.051$\pm$0.128 & 0.639$\pm$0.116 & -1.716$\pm$1.196 & 3.167$\pm$0.108 & 1.216$\pm$0.381 \\
\enddata

\end{deluxetable*}

\begin{figure}[htbp!]
\gridline{
   \fig{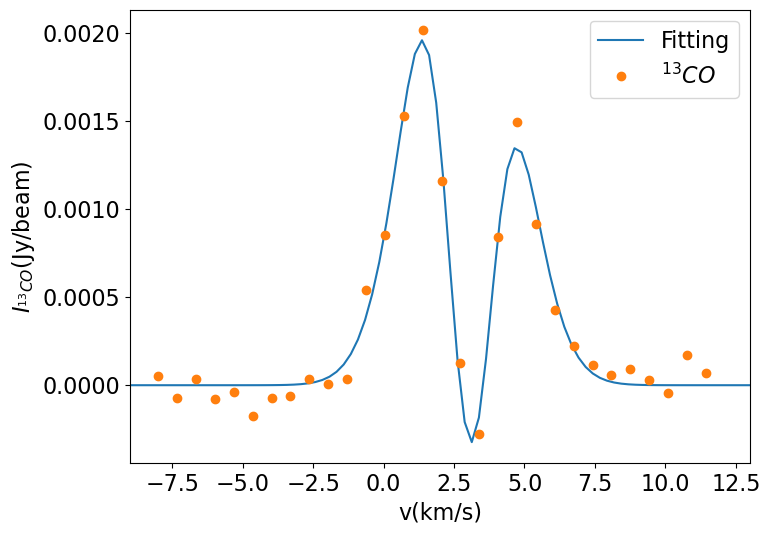}{0.45\textwidth}{a}
   \label{fig:figco2a}
    }
\gridline{
    \fig{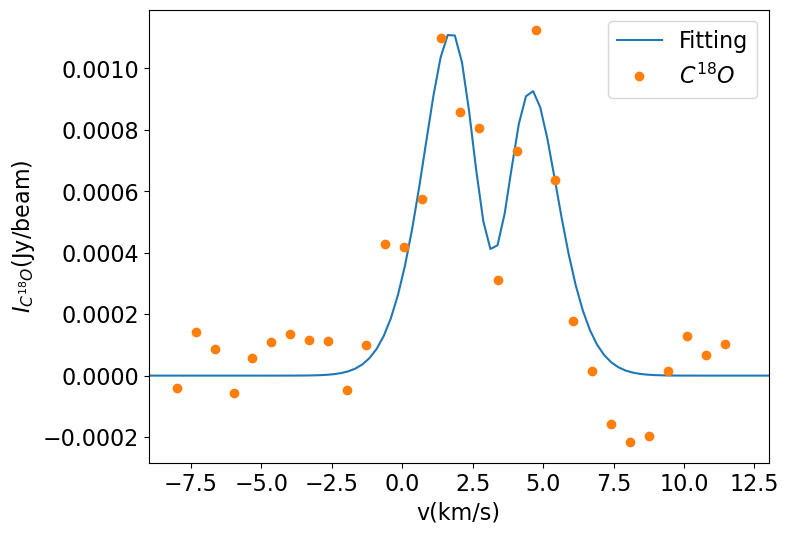}{0.45\textwidth}{b}
    \label{fig:figco2b}
    \fig{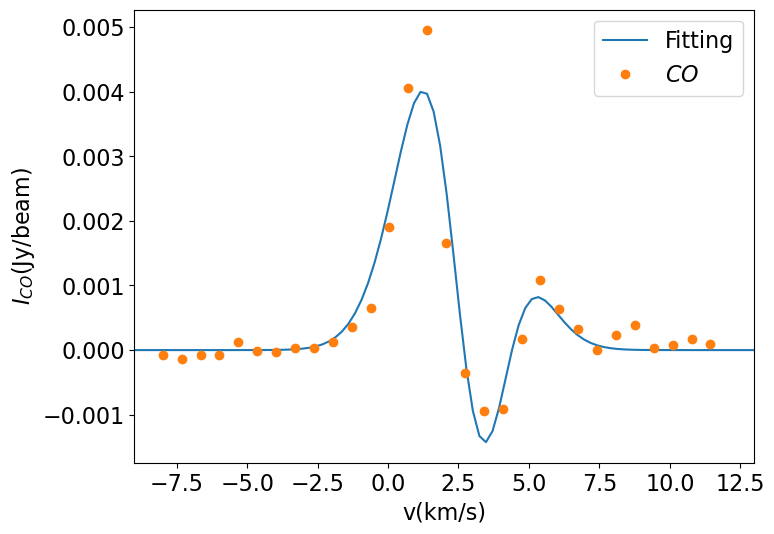}{0.45\textwidth}{c}
    \label{fig:figco2c}
	}
	\caption{CO line fitting images of different isotopologue emissions. }
\label{fig:figco2}
\end{figure}

\begin{figure}[htbp!]
\gridline{
   \fig{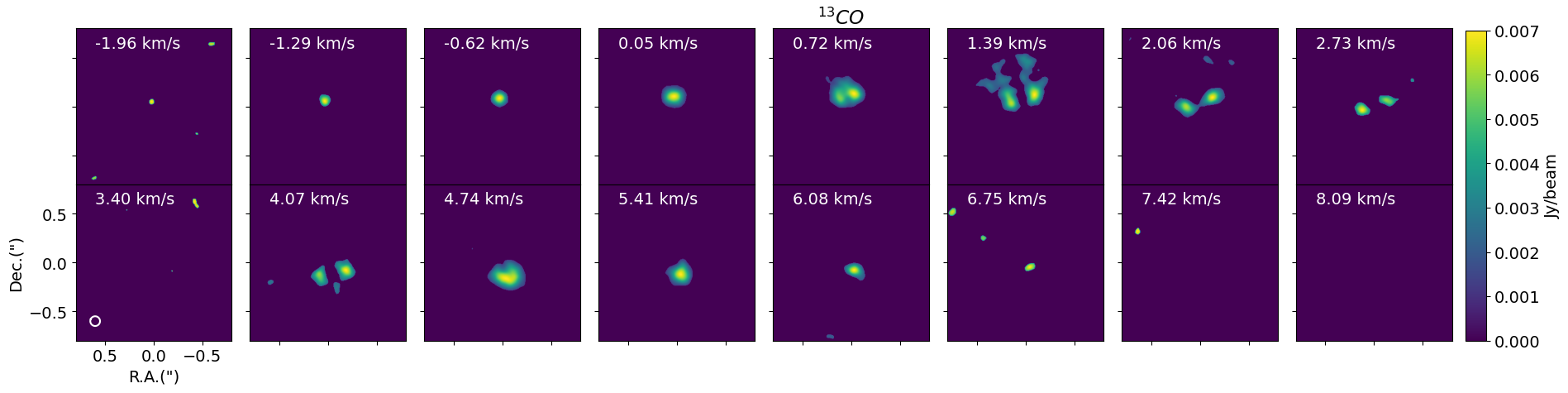}{0.9\textwidth}{a}
   \label{fig:figchannelmapa}
    }
\gridline{
    \fig{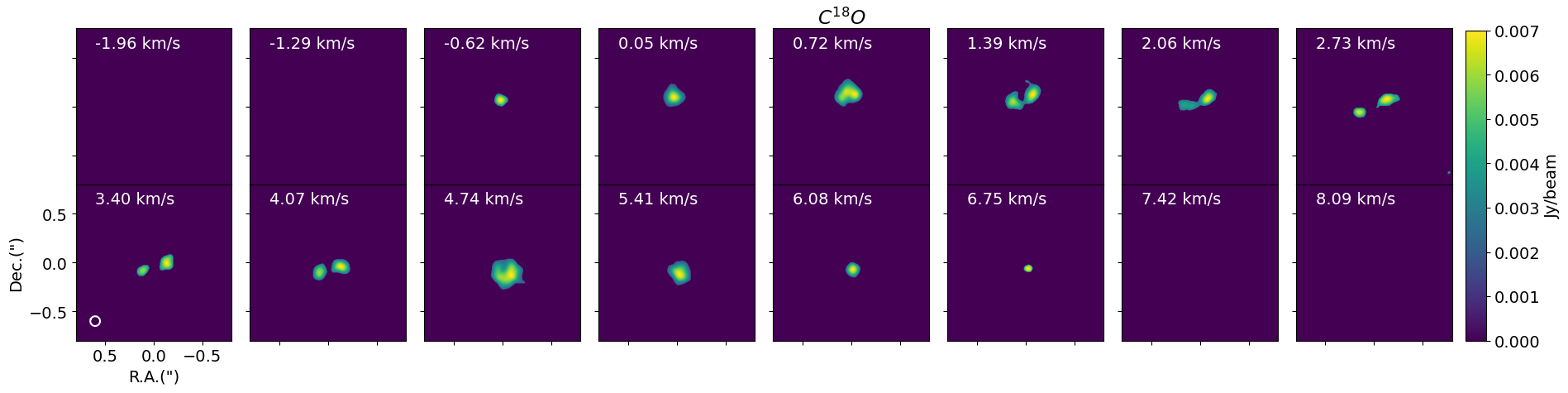}{0.9\textwidth}{b}
    \label{fig:figchannelmapb}
    }
\gridline{
    \fig{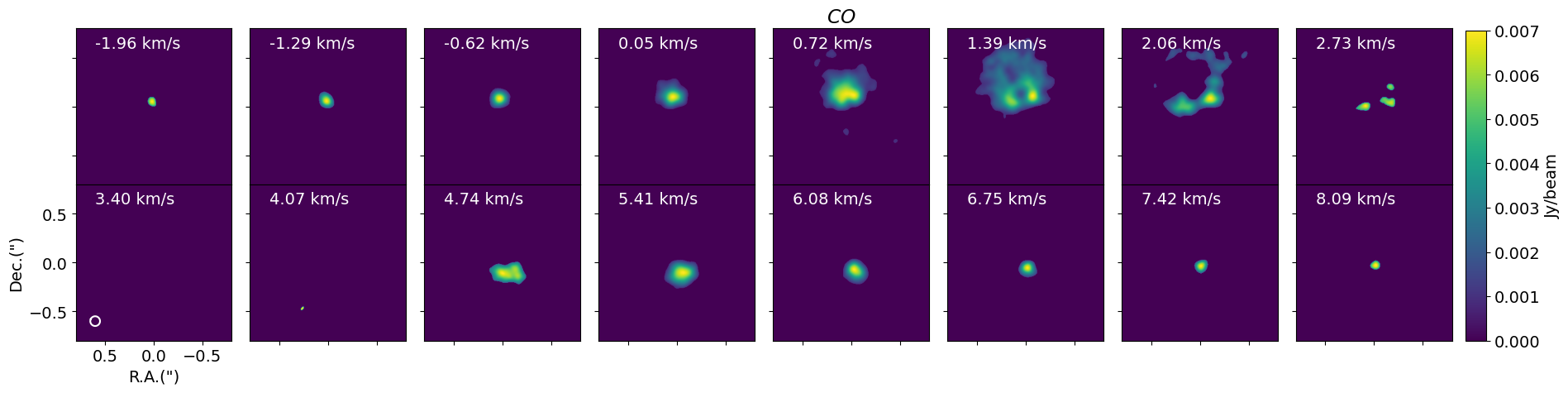}{0.9\textwidth}{c}
    \label{fig:figchannelmapc}
	}
	\caption{Channel maps of CO isotopologue emissions. Pixels fainter than 0.004 Jy/beam were blocked. }
\label{fig:figchannelmap}
\end{figure}

The fitting is done by using SciPy \textit{curve\_fit} command, with Trust Region Reflective algorithm. Since the disk is nearly face-on, we used intensities within $0.\arcsec5$ for the fitting. The result is shown in Figure~\ref{fig:figco2} and summarized in Table~\ref{tab:tab1}. The fitting results show that the emission lines center at Kinematic Local Standard of Rest (LSRK) velocity about 2.8 km/s, corresponding to barycentric velocity about -7.7 km/s, which is consistent with the radial velocity -7.3$\pm$1.64 km/s calibrated by GAIA \citep{GAIA2018}. The absorption features center at about 3.1 km/s, corresponding to a barycentric velocity of about -7.4 km/s resulting from foreground absorption. This absorption can be seen in the channel maps, especially for $^{13}$CO and CO line emissions (Figure~\ref{fig:figchannelmap}). According to the barycentric velocities, the absorption body is moving toward the disk. Besides, it should be noted that from the channel maps, the north structures (e.g., the 1.39km/s and 2.06km/s channels of $^{12}$CO map) are more extended than the south structures (e.g., the 5.4km/s and 6.08 km/s channels of $^{12}$CO map), this may indicate the gas structures in the disk are asymmetric.

\begin{figure}[htbp!]
\gridline{
    \fig{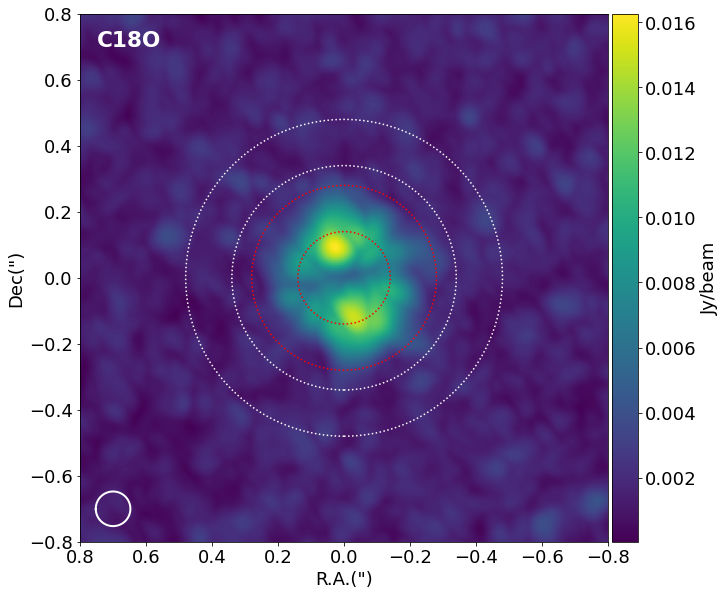}{0.5\textwidth}{a}
    \label{fig:figco3a}
    \fig{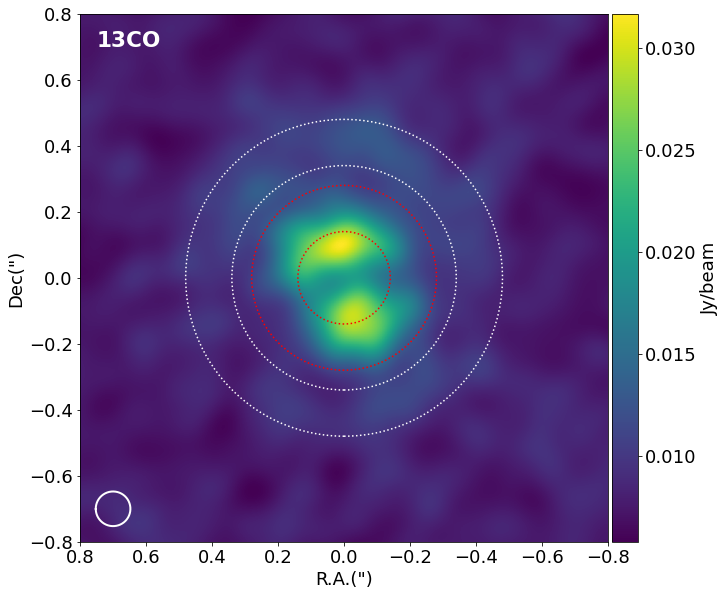}{0.5\textwidth}{b}
    \label{fig:figco3b}
	}
\caption{C$^{18}$O moment 8 image (a) and $^{13}$CO moment 8 image with continuum (b).}
\label{fig:figco3}
\end{figure}

Then we take some steps to estimate the Toomre Q parameters \citep{1964ApJ...139.1217T} to help understand the gravitational stability status inside the inner disk ($<$0.$\arcsec$28). For calculation, we unite the beam sizes of CO isotopologue images to $0.\arcsec105\times0.\arcsec105$ using the \texttt{imsmooth} command. Firstly, we make the moment 8 map of $C^{18}O$ (Figure~\ref{fig:figco3}(a)), which shows the peak intensities of every pixel. The CO isotopologue abundance ratio of the interstellar medium is $^{13}CO:^{12}CO=1:69$ and $C^{18}O:^{12}CO=1:570$ (i.e., $C^{18}O:^{13}CO\approx1:8$), as for protoplanetary disks, the abundance ratio in the protoplanetary disk of HD 163296 is $^{13}CO:^{12}CO=1:67$ and $C^{18}O:^{13}CO=1:7$ \citep{Qi2011-fb} and \citet{Yoshida2022-jd} founded out that in the disk around TW Hya, the $^{13}CO:^{12}CO=1:21$ at disk radius of 70-110 AU and $^{13}CO:^{12}CO=1:84$ for disk radius larger than 130 AU. These results are all significantly smaller than the ratios in our observations. Therefore, the $^{12}CO$ and $^{13}CO$ lines are optically thick, and we can use their brightness temperatures as gas temperatures. Here we use the moment 8 map of $^{13}$CO with continuum (Figure~\ref{fig:figco3}(b)) to calculate the brightness temperature $T_{^{13}CO}$ using the Planck function $I_{^{13}CO,peak}=B(\nu_{^{13}CO},T_{^{13}CO})$. Then the peak optical depth of C$^{18}$O emission $\tau_{C^{18}O, peak}$ can be derived via the radiative transfer equation:

\begin{equation}
    I_{C^{18}O,peak}=B(\nu_{C^{18}O},T_{^{13}CO})(1-\mbox{exp}(-\tau_{C^{18}O,peak}))
\end{equation}

After that, we derive the $C^{18}O$ column density $N_{C^{18}O}$ using the equation given by \citet{Mangum2015-ud}:

\begin{equation}
    N_{C^{18}O}=\frac{3h}{8\pi^3\mu^2J_u}(\frac{kT_{ex}}{hB_0}+1/3)\mbox{exp}(\frac{E_{J_u}}{kT_{ex}})[\mbox{exp}(\frac{h\nu}{kT_{ex}})-1]^{-1}\int \tau dv
\end{equation}

Here $\mu$, $J_u$, and $B_0$ are the dipole moment, the rotational quantum number of the line transition upper level, and the rigid rotor rotational constant, respectively. The excitation temperature $T_{ex}$ equals the gas temperature. We assume that the line optical depth has a Gaussian profile with the peak value of $\tau_{peak}$ and the width of $\Delta v$ then $\int \tau dv\approx\sqrt{2\pi}\tau_{peak}\Delta v$, and $\Delta v=2\sqrt{2 \mbox{ln} 2}\Delta V_{emis}$. After knowing the column density, the gas (hydrogen) surface density $\Sigma_g$ can be calculated via:
\begin{equation}
    \Sigma_g=m_{H_2}N_{C^{18}O}/\chi
\end{equation}
here $\chi$ is the ratio of $C^{18}O$ to $H_2$, and we use the interstellar abundance 1.79$\times10^{-7}$ \citep{1999RPPh...62..143W}. Assuming that the disk surface density $\Sigma$ is dominated by the gas surface density, i.e., $\Sigma\approx\Sigma_g$, then the Toomre Q parameter can be derived from:
\begin{equation}
    Q=\frac{c_k\Omega_k}{\pi G \Sigma_g}
\end{equation}

\begin{figure}[htbp!]
\gridline{
   \fig{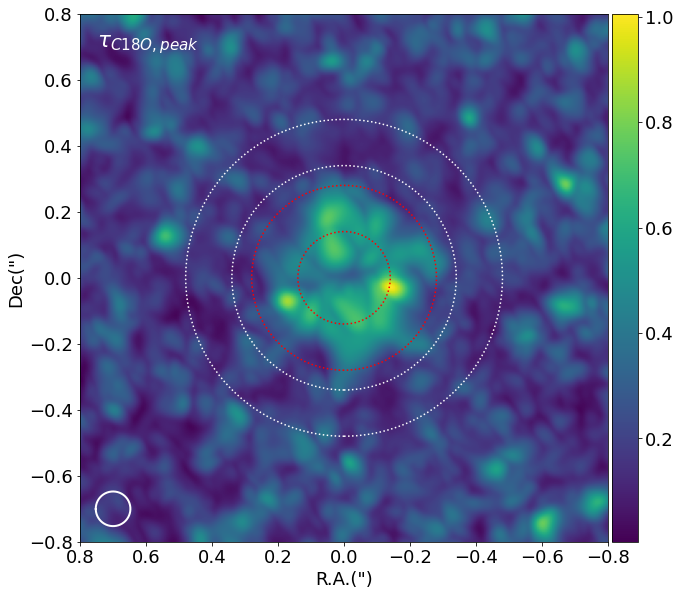}{0.45\textwidth}{a}
   \label{fig:figco4a}
    }
\gridline{
    \fig{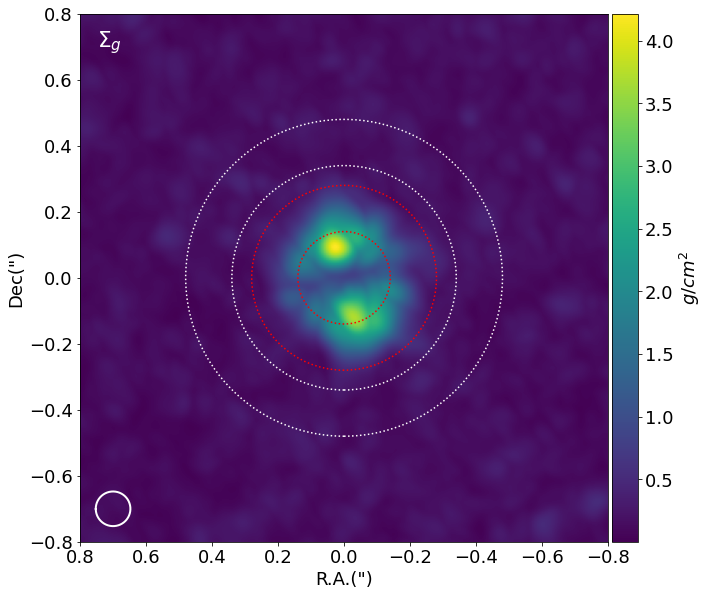}{0.45\textwidth}{b}
    \label{fig:figco4b}
    \fig{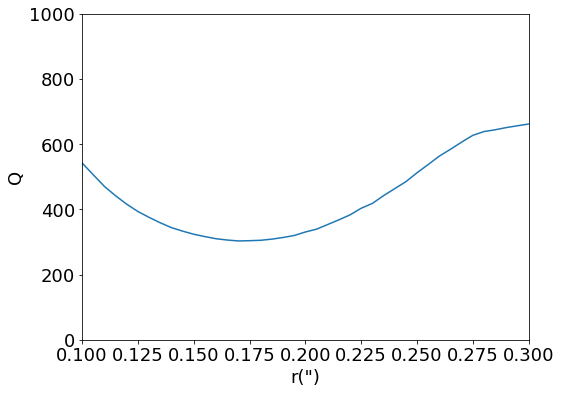}{0.45\textwidth}{c}
    \label{fig:figco4c}
	}
	\caption{(a): $\tau_{C^{18}O,peak}$ image; (b): gas surface density image; (c): radial profile of Toomre Q parameters;}
\label{fig:figco4}
\end{figure}

Here $c_k=\sqrt{kT/m_{H_2}}$ is the sound speed, and $\Omega_k$ is the epicyclic frequency, which we assume that it equals the angular velocity, i.e., $\Omega_k=\sqrt{GM_{star}/r^3}$. The result is shown in Figure~\ref{fig:figco4}(c), considering that Q$\gg$1, the inner disk is gravitational stable.

There are some caveats in this calculation, though. Firstly, emission from these three tracers CO, $^{13}$CO, C$^{18}$O is likely coming from different vertical heights in the disc, thus using $^{13}$CO and C$^{18}$O together to calculate $I_{C^{18}O, peak}$ assumes the two are tracing similar parts of the disc, which is not necessarily true. Secondly, we assume that the ratio of $C^{18}O$ to $H_2$ is the same as the interstellar abundance, which is also not necessarily true: the CO molecules suffer from several effects such as CO freeze-out \citep[e.g.,][]{Qi2011-fb}, isotope-selective photodissociation \citep[e.g.,][]{Miotello2016-gw,Miotello2018-lt} and CO depletion \citep[e.g.,][]{Krijt2020-oe}, these effects will lower the observed abundance of CO molecules; thus the actual disk mass could be higher, and Q parameter could be smaller than our calculation. 

\section{Discussions} \label{sec:Discussions}

\subsection{Disk Morphology}

The two asymmetric structures in the outer ring are quite elongated, making them the elongated vortices suggested in previous studies, which could be caused by dead zone \citep[e.g.,][]{Regaly2012-zz} or a slow-growing planet \citep[e.g.,][]{Hammer2017-dq,Hammer2019}. \citet{Hammer2019} suggested that compared to dead zone-induced elongated vortices, planet-induced elongated vortices frequently have off-center peaks due to repeated perturbations from the planet’s spiral arms. In the outer ring, two peaks of the asymmetric structures (the North Arc and the South Arc) exist. Their azimuthal profiles are both asymmetric, consistent with the characteristics of planet-induced vortices suggested by \citet{Hammer2019}. Therefore, we suggested that the two asymmetric structures in the outer ring are induced by a growing planet between the rings rather than a dead zone. Furthermore, \citet{Hammer2019} pointed out that such elongated asymmetric structures may also be accompanied by trojan dust that is co-orbital with the planet at the L5 Lagrange point, and the azimuthal profile could be changed by the interaction
between the planet’s spiral density waves and its associated vortex. Our sub-millimeter observations do not catch obvious structures which could be related to trojan dust, while \citet{Muro-Arena2020} caught one "kink" structure between the gap, indicating a planet there. Here we suggested that this "kink" may also be the trojan dust induced by the planet, and further study is needed to confirm this. As for the latter, \citet{Hammer2019} suggested the change could happen for a planet at Jupiter’s separation from the Sun in a few years. Considering the planet of SR 21 is farther from the star and less massive (see Section 4.3), it may take more time for the azimuthal profile to change.

Despite \citet{Hammer2019}, \citet{Li2020-pt} suggested that one original large asymmetric structure will break into small clumps due to the feedback of formed dust onto the gas, and the broken clumps can be more prominent at a longer wavelength, which could help explain the multiple clumps observed in Band 3. However, as discussed in Section 3.1.1, the clump structures may not be real; future observations are needed to confirm them.

In addition, previous research like \citet{Baruteau2015-gj} suggested that the peaks of asymmetric structures in different dust sizes will have offsets if they are vortices triggered by the RWI; thus the peaks of the asymmetric structures will be different in different wavelengths. \citet{Cazzoletti2018-gi} confirmed offsets of the asymmetric structures in the disk around HD 135344B between different wavelengths, but the offset is in the opposite direction of the theoretical prediction. As for SR 21, the peak of North Arc in Band 3 is $300^\circ\pm5^\circ$, while in Band 6, it is $305^\circ\pm5^\circ$, so in Band 6, the peak is $5^\circ\pm5^\circ$ ahead of Band 3. The South Arc has a similar case: in Band 6, the peak is $5^\circ\pm5^\circ$ ahead of Band 3 (170$^\circ\pm5^\circ$ vs. 165$^\circ\pm5^\circ$). While for the Inner Arc, the peak in Band 6 is $5^\circ\pm5^\circ$ below the peak in Band 3 (165$^\circ\pm5^\circ$ vs. 170$^\circ\pm5^\circ$). The SR 21 disk is nearly face-on, so it is hard to judge which side is nearer to us, then it is hard to know whether the gas in the disk rotates clockwise or counterclockwise. Therefore it is hard to judge if the observed offsets are consistent with the theoretical prediction. The peak offset of the asymmetric structure in HD 135344B between Band 3 and Band 4 is about 25$^\circ$, and about 40$^\circ$ between Band 3 and Band 7. Compared to this, the peak offsets in SR 21's asymmetric structures are pretty smaller than HD 135344B. One possible explanation for this is that the asymmetric structures in SR 21 are newly formed, and it may still need some time to develop a significant peak shift as HD 135344B.

\subsection{Dust Size}

We derived the spectral indices map using the Band 3 and 6 data to estimate the maximum dust size ($a_{max}$). In this step, we re-image the Band 6 data to make sure it has the same beam size (0.$\arcsec$10$\times$0.$\arcsec$09, and PA=90$^\circ$) and UV-distance ($50-5000$ k$\lambda$) as the Band 3 data, and register the two images based on the proper motion data got by \citet{GAIA2018}. The spectral indices $\alpha$ were calculated by: 

\begin{equation}
    \alpha=\frac{log(L_{3}/L_{6})}{log(\nu_{3}/\nu_{6})}
    \label{spectral_index}
\end{equation}

Here $L_{3}$ and $L_{6}$ are the intensities of Band 3 and Band 6 images. $\nu_{3}$=110.2 GHz and $\nu_{6}$=217.0 GHz are the rest frequencies of Band 3 and 6 observations, respectively. Only the pixels brighter than 4$\sigma$ are used to calculate the index. In addition, the errors $\Delta\alpha$ were given by:

\begin{equation}
    \Delta\alpha=\sqrt{(\frac{\partial\alpha}{\partial L_{3}})^2\Delta L_{3}^2+(\frac{\partial\alpha}{\partial L_{6}})^2\Delta L_{6}^2}
    \label{spectral_index_error}
\end{equation}

\begin{figure}[htbp!]
\gridline{
   \fig{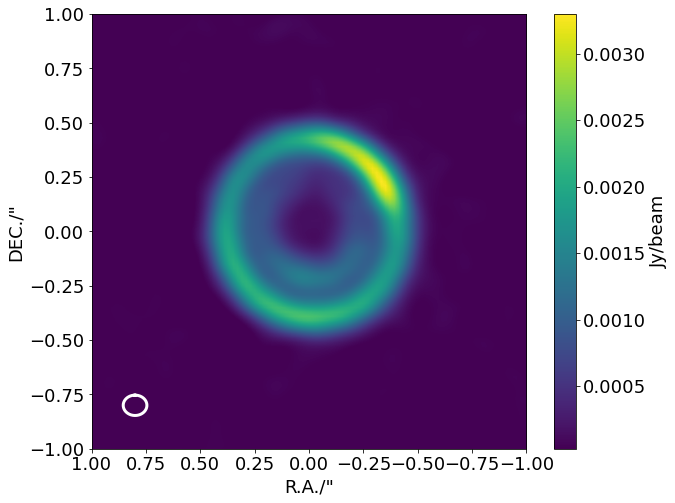}{0.4\textwidth}{a}
   \label{fig:fig5a}
    }
\gridline{
    \fig{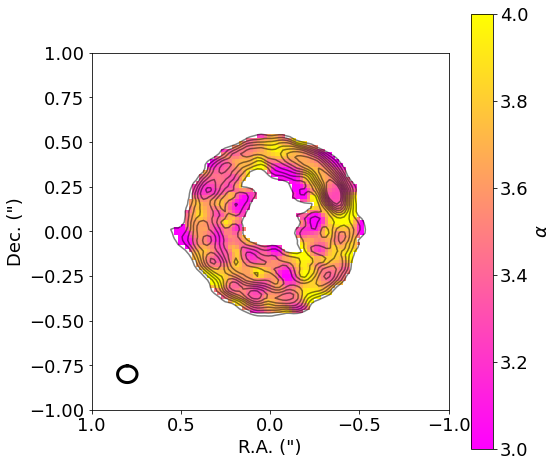}{0.3\textwidth}{b}
    \label{fig:fig5b}
    \fig{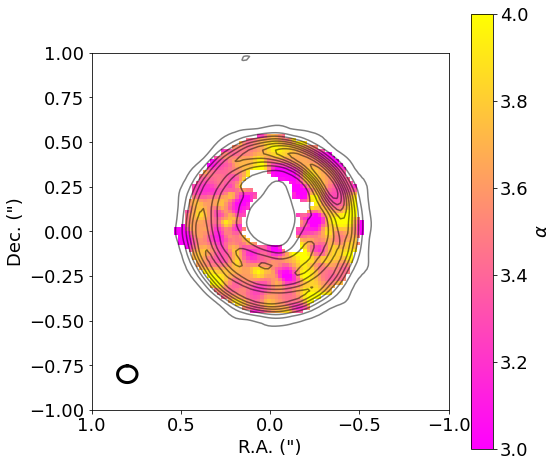}{0.3\textwidth}{c}
    \label{fig:fig5c}
    \fig{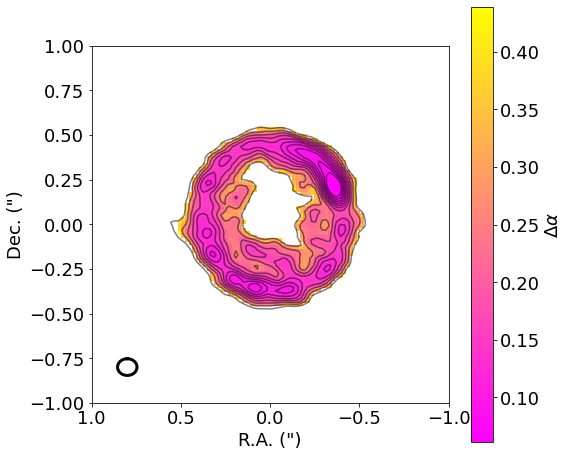}{0.3\textwidth}{d}
	\label{fig:fig5d}
	}
\gridline{
    \fig{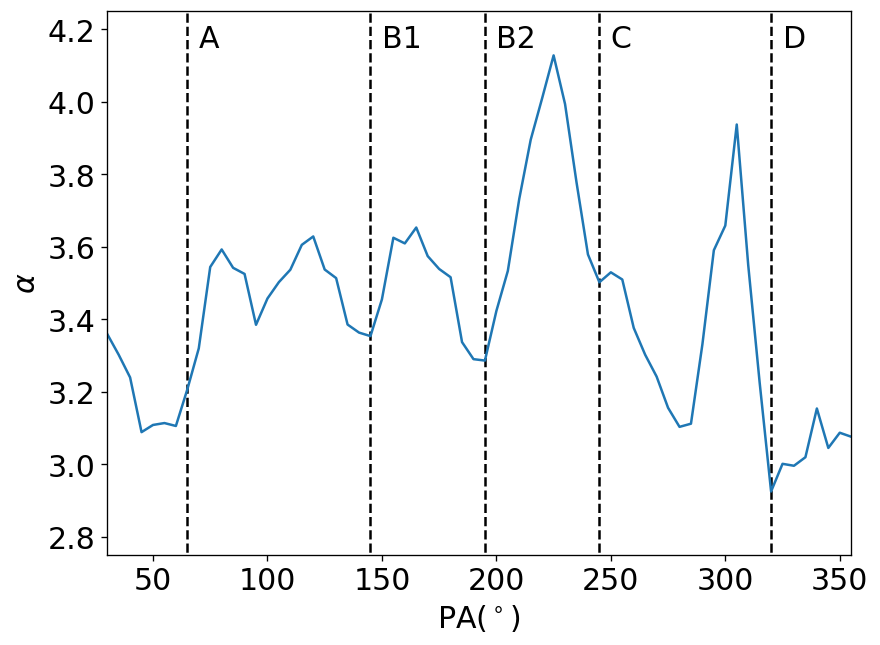}{0.45\textwidth}{d}
    \label{fig:fig5b}
    \fig{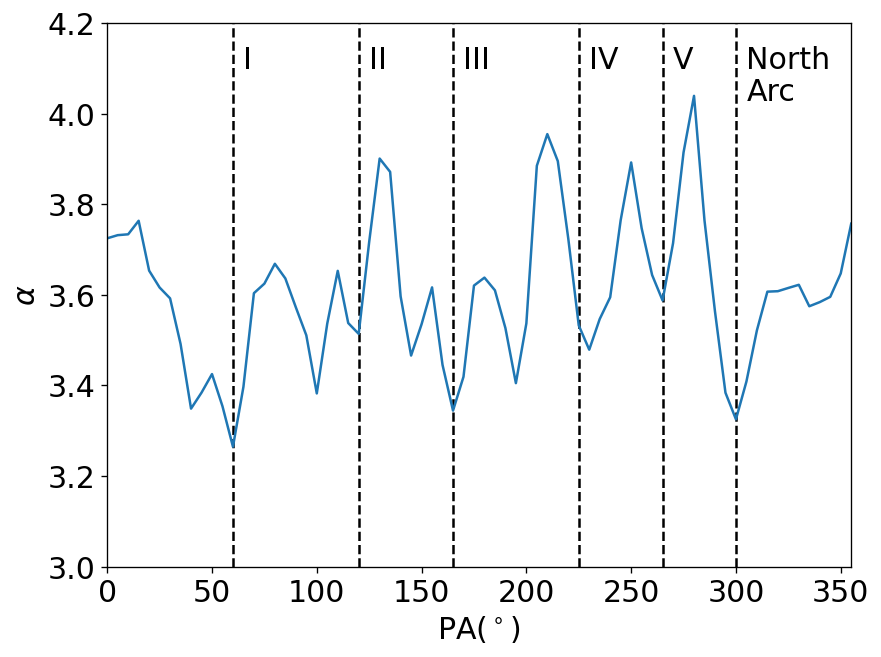}{0.45\textwidth}{e}
    \label{fig:fig5c}
	}
	\caption{(a): SR 21 Band 6 image with the same beam and UV range as Band 3; (b) and (c): Spectral Index map of SR 21 derived from Band 3 and Band 6 images, with the same beam size and UV-distance. The contours in Figure (b) show Band 3 data, starting from 5.6$\times10^{-5}$ Jy/beam with contour interval $\sim1.5\times10^{-5}$ Jy/beam; the contours in Figure (c) corresponds to the Band 6 data shown in (a), starting from 8$\times10^{-5}$ Jy/beam with contour interval $\sim3.2\times10^{-4}$ Jy/beam. (d): Errors of spectral indices, overplotted with Band 3 Contours; (e): the azimuthal profile of the inner ring indices, the dashed lines show the peaks detected in Band 3; (f): the azimuthal profile of the outer ring indices, the dashed lines show the clumps detected in Band 3.}
\label{fig:fig5}
\end{figure}

Here $\Delta L_{3}=1.4\times10^{-5}$Jy/beam and $\Delta L_{6}=3\times10^{-5}$Jy/beam are the RMS noise level of Band 3 and 6 images used here. The results are shown in Figure~\ref{fig:fig5}. From the figures, one can see that the spectral indices of the disk are mainly between 3.0 and 4.0. At the clumps, the spectral indices are generally smaller than 3.5, smaller than other places, which indicates larger grain sizes at the locations of the clumps. This can be seen more clearly in the azimuthal profiles of the inner ring (Figure~\ref{fig:fig5}(e)) and the outer ring (Figure~\ref{fig:fig5}(f)): the lowest spectral index in the inner ring is at clump D, about 3.0, while the lowest spectral index in the outer ring is at clump I, about 3.3. Note that for the azimuthal profile of the inner ring, the values between PA 0-30$^\circ$ are not drawn because the flux in Band 3 here is smaller than 4$\sigma$, and the Band 3 flux around PA 280$^\circ$ is also smaller than 4 $\sigma$, so one can see there is one drop at this position angle. Based on the simulations like \citet{Birnstiel2018-jb}, spectral indices $\sim$3-4 generally correspond to $cm$ size of $a_{max}$ when q=3.5 is assumed in the dust size distribution $n(a)\propto a^{-q}$.

Estimating $a_{max}$ from spectral indices is not accurate when the disk is optically thick and/or dust scattering cannot be neglected (\citealt{Liu2019ApJ...877L..22L}). Therefore, we used the azimuthal profiles of the inner ring and the outer ring in Band 3 and 6 to perform fittings, hoping to understand the dust distributions of the rings. We use the method developed by \citet{Carrasco-Gonzalez2019-mq} and \citet{Sierra2020-zs}. When the dust scattering is taken into account, the intensity of the disk can be written as:

\begin{equation}
    I_\nu=B_\nu(T)[(1-\mathrm{exp}(-\tau_\nu/\mu))+\omega_\nu F(\tau_\nu,\omega_\nu)]
    \label{intensity}
\end{equation}

here $\tau_\nu=\Sigma_d (\kappa_\nu+\sigma_\nu)$ is the optical depth. $\Sigma_d$ is the dust surface density, $\kappa_\nu$, and $\sigma_\nu$ represents the absorption and scattering coefficient respectively. Note that for the scattering coefficient, we use the effective scattering coefficient defined as $\sigma^{eff}_{\nu}=(1-g_\nu)\sigma_\nu$, here $g_\nu$ is the asymmetry parameter. $\omega_\nu=\sigma_\nu/(\kappa_\nu+\sigma_\nu)$ is the dust albedo, $\mu=cos(i)$ is the cosine of the inclination angle $i=15^\circ$. $F(\tau_\nu,\omega_\nu)$ is defined as:

\begin{equation}
    F(\tau_\nu,\omega_\nu)=\frac{1}{\mathrm{exp}(-\sqrt{3}\epsilon_\nu \tau_\nu)(\epsilon_\nu-1)-(\epsilon_\nu+1)}\times[\frac{1-\mathrm{exp}(-(\sqrt{3}\epsilon_\nu+1/\mu)\tau_\nu)}{\sqrt{3}\epsilon_\nu\mu+1}+\frac{\mathrm{exp}(-\tau_\nu/\mu)-\mathrm{exp}(-\sqrt{3}\epsilon_\nu\tau_\nu)}{\sqrt{3}\epsilon_\nu\mu-1}]
\end{equation}

here $\epsilon_\nu=\sqrt{1-\omega_\nu}$. We assumed q=3.5 in our fitting, then equation~\ref{intensity} will depend on only three parameters: $T_d$, $\Sigma_d$ and $a_{max}$. And since we only have two bands, we fixed the dust temperature at the inner ring (at 0.$\arcsec$21) and the outer ring (at 0.$\arcsec$41) so that there are only two free parameters. By assuming that the temperature of grown dust is equal to the disk mid-plane temperature, we calculated the dust temperature using Equation (3) in \citet{Huang2018-sa}, which approximates the mid-plane temperature profile of a passively heated, flared disk in radiative equilibrium. The DSHARP dust optical constants (\citealt{Birnstiel2018-jb}) are employed in our fitting. In details, we explored a series of $\Sigma_d$ and $a_{max}$ values, calculating the flux $I_{\nu,model}$ in different wavelengths using equation~\ref{intensity}, then derived the probability via the following equation:

\begin{equation}
    p\propto\mathrm{exp}(-\sum_{\nu}(I_{\nu}-I_{\nu,model})/2\sigma_\nu^2)
\end{equation}

Here $\sigma_\nu$ is the standard deviation of the azimuthal profiles. The explored parameter space is a logarithmically spaced grid between $10\mu m\leqslant a_{max} \leqslant 1 cm$ and $0.001 g/cm^2 \leqslant \Sigma_d \leqslant 10 g/cm^2$, each divided into 1000 bins.

\begin{figure}[htbp!]
\gridline{
    \fig{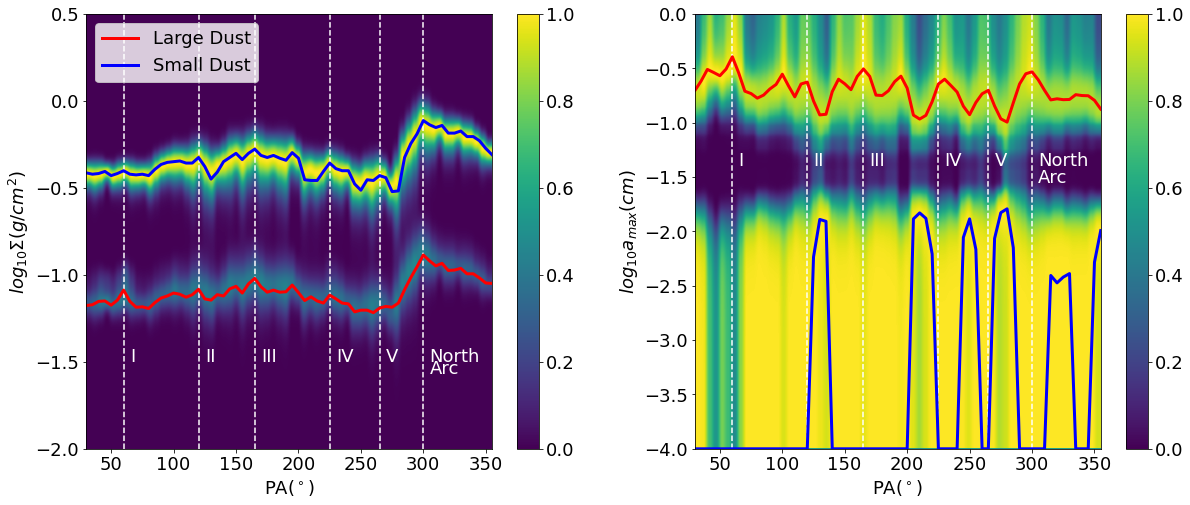}{0.5\textwidth}{a}
    \label{fig:fig6a}
    \fig{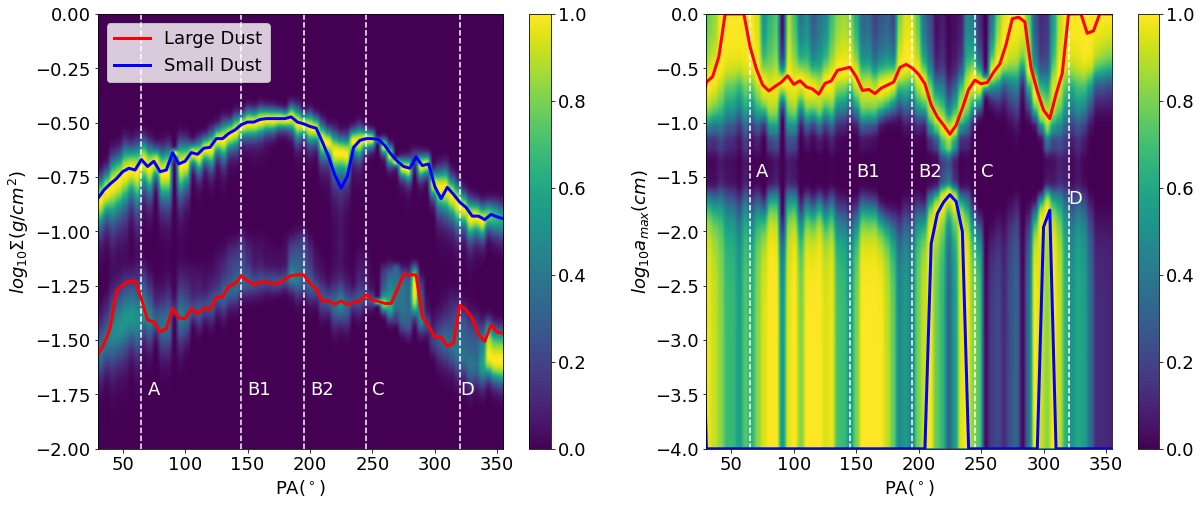}{0.5\textwidth}{b}
    \label{fig:fig6b}
	}
\caption{Fitting results for the surface density and maximum dust size of the outer ring (a) and the inner ring (b). The color map shows the normalized probability, red and blue lines show the best solutions for the large and small dust, respectively. The white dashed lines show the clumps detected in Band 3.}
\label{fig:fig6}
\end{figure}

The fitting result is shown in Figure~\ref{fig:fig6}. At each position, the fittings generally lead to two families of probable solutions, one with $a_{max}\lesssim$300 $\mu m$ and a higher surface density (small dust solution), and the other with $a_{max}\gtrsim1 mm$ and lower surface density (large dust solution). Such a degeneracy in the SED fittings is also pointed out by the recent case studies on the CW Tau disk (\citealt{Ueda2022arXiv220316236U}). We also draw the positions of the clumps in the images (dashed line). Generally, for the large dust solutions, the clumps correspond to larger solutions than other places, while for the small dust solutions, the clumps correspond to smaller solutions.

Theoretical studies favor the large dust solutions at the peaks of asymmetric structures and clumps; however, from the observational view, it is hard to judge which solution is proper. The actual case could be a mixture of these two solutions: in some places, the large-dust solution is proper, while in other places, the small-dust solution is proper. Future observations from longer wavelengths or polarization observations can help distinguish these possibilities.

\subsection{Potential Planet}

\citet{Muro-Arena2020} suggested that a potential planet could cause the gap between the inner and outer rings. They suggested that the planet favors a low mass ($\sim 1M_{Jup}$) based on the simulation results of \citet{De_Juan_Ovelar2013-zg} after comparing the ring location in the near-infrared band and sub-millimeter band. Keck L-band (3.426–4.126 $\mu$m) observations have excluded companions more massive than 13 $M_{Jup}$. In the results of \citet{De_Juan_Ovelar2013-zg}, the simulated sub-millimeter image of the disk does not show a bright inner ring, which is different from the observation. Therefore, based on our observation data, we consider re-constraining the potential planet mass. In our Band 6 data, the gap size is about 0.$\arcsec$06, and its center is about 0.$\arcsec$31, so we tried to constrain the potential planet mass from the disk gap sizes. From our observation result, it is hard to judge whether the planet opens a gas gap, so that we will discuss them separately. If the planet only opens a dust gap, the gap size will be proportional to the planet Hill radius $R_H$ \citep[e.g.,][]{Lodato2019-nq}:

\begin{equation}
    R_H=(\frac{M_p}{3M_*})^{1/3}R_{gap}
    \label{Hill_radius}
\end{equation}

and the planet's mass can be derived by:

\begin{equation}
    M_{p,dust}(R_{gap})=M_*(\frac{\Delta_{gap}}{CR_{gap}})^3
    \label{dust_gap}
\end{equation}

here $\Delta_{gap}$ and $R_{gap}$ represent the width and location (radius) of the gap, respectively. $4.5<C<7$ is the proportional constant \citep{Rosotti2016-zs}. Based on this, we can estimate the planet which opens the gap should be about 12-44$M_\oplus$.

As for the case that the planet opens a dust gap and a gas gap, we assume that the gas gap is equal to the dust gap, then according to the empirical equation given by \citet{Kanagawa2016-cf}, the planet mass should be:

\begin{equation}
    M_{p,gas}(R_{gap})=2.1\times10^{-3}M_*(\frac{\Delta_{gap}}{R_{gap}})^2(\frac{h_{gap}/R_{gap}}{0.05})^{3/2}(\frac{\alpha}{10^{-3}})^{1/2}
    \label{gas_gap}
\end{equation}

here $h_{gap}$ and $\alpha$ represent the scale height and viscosity at the gap, respectively. The aspect ratio $h_{gap}/R_{gap}$ can be derived as:

\begin{equation}
    h_{gap}/R_{gap}=c_k/\Omega
    \label{aspect_ratio}
\end{equation}

here $c_k$ is the sound speed and $\Omega$ is the angular velocity. Using Equation (3) in \citet{Huang2018-sa}, the temperature at $0.\arcsec31$ is about 31 K, and then we can derive the aspect ratio here is about 0.06. As for the viscosity $\alpha$, since the asymmetric structure we detected could be the result of RWI vortices, which requires low viscosity ($\alpha\lesssim10^{-4}$) \citep{Regaly2012-zz} for the vortices to survive. So we assume  $\alpha\sim10^{-4}-10^{-3}$, and the result will be about 18-58 $M_\oplus$ (0.06-0.18 $M_{Jup}$). 

The criteria to judge if the planet opens a gas gap is to see if the planet reaches pebble isolation mass \citep{Morbidelli2012-dw,Lambrechts2014-fi,Birnstiel2018-jb}. Planet mass below pebble isolation mass could not or can only open a shallow gas gap. According to equations 10 and 11 in \citet{Birnstiel2018-jb}, the pebble isolation mass for SR 21 could be about 30-50 $M_\oplus$; thus, one gas gap opened by the planet may exist. Confirming the gas gap size using sub-millimeter observations in the future could help better constrain the planet's mass. 

Planets excite density waves, which may look like spiral arms in scattered light \citep{Dong2015-bh}. However, the density waves excited by a sub-Saturn mass planet are probably too weak to be visible in current observations, which typically require (multi-)Jupiter mass planets \citep{Dong2017-ps}. In addition, if the two spirals identified in \citet{Muro-Arena2020} are produced by one planet, their large azimuthal separation suggests a multi-$M_{Jup}$ planet too \citep{Fung15}. Thus, the predicted gap-opening planet cannot explain the spiral arm-like features seen in \citet{Muro-Arena2020}. While the gravitational instability may also produce spiral arms in scattered light \citep{dong15giarm}, it is unlikely the cause of the spirals in this case (Section 3.2). 

We propose that the most prominent spiral arm-like feature identified in the SPHERE observation, Spiral 1 (Figure~\ref{fig:fig4}), could be the scattered light counterpart of the Inner Arc, which itself is a dust-trapping vortex. As shown in \citet{Marr22}, vortices appear as arcs in scattered light, which may look like spirals in inclined disks. The excellent position coincidence between Spiral 1 and Inner Arc supports this hypothesis. In addition, vortices can cast shadows in the outer ring due to their elevated surface height. Meanwhile, the region outside Spiral 1 in scattered light appears fainter than other regions at the same radius, matching the shadow prediction as well. This could help explain why the spiral arm in SR 21 looks so different from other targets.

There are some caveats in this estimation, though. Firstly, the gap may be triggered by just one planet inside the central cavity. In the simulation work done by \citet{Gomes2015-qy}, they showed that one planet inside the central cavity of the transitional disk could successfully induce two vortices and the gap between them. However, considering the planet's orbit is hard to constrain, it is hard to estimate its mass under this scenario. Secondly, our calculation is aspect-ratio dependent, and the aspect-ratio is dependent on the temperature of the disk middle plane. The result will also change if the actual temperature deviates from Equation (3) in \citet{Huang2018-sa}. The estimation result can be improved by a better understanding the temperature profile.

\section{Conclusions} \label{sec:cite}

The protoplanetary disk around SR 21 holds many characteristics that could help us understand disk evolution and planet formation. By combining observation results in different wavelengths, the main conclusions are listed below:

1. Based on the ALMA Band 3 and Band 6 dust continuum images, the protoplanetary disk around SR 21 under sub-millimeter bands consists of two rings with a radius of about 0.$\arcsec$14-0.$\arcsec$28 and 0.$\arcsec$34-0.$\arcsec$48. Three crescent-like asymmetric structures, one in the inner ring and two in the outer ring, are detected, while CO isotopologues images of Band 6 show that the emissions are mainly inside the inner ring ($\leqslant0.\arcsec28$), which could be due to the foreground absorption; the $^{12}$CO gas structure in the north is more extended than the south, this may indicate that the gas distribution in the disk is asymmetric; 

2. The asymmetric structures in the outer ring are elongated, and their azimuthal profiles are asymmetric, and in Band 3 data, some small clumps in the rings may exist. Features like off-center peaks of the asymmetric structures indicate that the asymmetric structures are elongated due to the interaction of a slow-growing planet rather than dead zone, as \citet{Hammer2017-dq, Hammer2019} suggested;

3. The fitting using Band 3 and Band 6 data suggests two solutions of the maximum dust sizes in the disk: small dust ($a_{max}\lesssim$300 $\mu m$) solution and large dust ($a_{max}\gtrsim 1 mm$) solution. Future observations from longer wavelengths or polarization observations can help distinguish them;

4. According to the gap size and the derived aspect ratio, the potential planet in the gap between the inner and outer rings should reach 18-58$M_\oplus$ corresponding to the viscosity $\alpha\sim10^{-4}-10^{-3}$. Planet at this mass is hard to trigger visible spiral arms in the near-infrared band, and we propose that the spiral arm-like features identified in the SPHERE observation could just be the scattered light counterpart of the Inner Arc.

\acknowledgments

We would like to thank the insightful and constructive comments from an anonymous referee that greatly improved the quality of this manuscript. This paper makes use of the following ALMA data: ADS\/JAO.ALMA\#2018.1.00689.S and \#2017.1.00884.S. ALMA is a partnership of ESO (representing its member states), NSF (USA), and NINS (Japan), together with NRC (Canada), MOST, and ASIAA (Taiwan), and KASI (Republic of Korea), in cooperation with the Republic of Chile. The Joint ALMA Observatory is operated by ESO, AUI/NRAO, and NAOJ. The National Radio Astronomy Observatory is a facility of the National Science Foundation operated under a cooperative agreement by Associated Universities, Inc. We also thank Prof. Pinilla, Paola for sharing her reduced ALMA Band 3 data, and Anibal Sierra for sharing his code and useful discussions. This work was supported by NAOJ ALMA Scientific Research Grant Numbers 2019-12A. M.T. is supported by JSPS KAKENHI grant Nos.18H05442,15H02063,and 22000005. Y.H. is supported by the Jet Propulsion Laboratory, California Institute of Technology, under a contract with the National Aeronautics and Space Administration (80NM0018D0004).

\vspace{5mm}
\facilities{ALMA}

\software{CASA}

\bibliography{sample63}{}
\bibliographystyle{aasjournal}

\end{document}